\documentstyle[12pt]{article}
\input{epsfig.sty}
\newcommand{\beq}{\begin{eqnarray}}
\newcommand{\eeq}{\end{eqnarray}}
\begin{document}
\title{Heavy Quark State Production In p-p Collisions}
\author{Leonard S. Kisslinger\\
Department of Physics, Carnegie Mellon University, Pittsburgh, PA 15213\\
Ming X. Liu and Patrick McGaughey\\
P-25, Physics Division, Los Alamos National Laboratory, Los Alamos, NM 87545}
\date{}
\maketitle
\begin{abstract}
  We estimate the relative probabilities of $\Psi'(2S)$ to $J/\Psi$ production 
at BNL-RHIC and $\Upsilon(nS)$ production at the LHC and Fermilab in p-p
collisions, using our recent theory of mixed heavy quark hybrids, in which the
$\Psi'(2S)$ and $\Upsilon(3S)$ mesons have approximately equal normal 
$q\bar{q}$ and hybrid $q\bar{q}g$ components.

\end{abstract}
\noindent
PACS Indices:12.38.Aw,13.60.Le,14.40.Lb,14.40Nd
\vspace{1mm}

\section{Introduction}

   There has been a great deal of interest in the production and polarization
of heavy quark states in proton-proton collisions.  This was motivated in
part by the $J/\Psi,\Psi'$ production anomaly\cite{cdf97}, in which the
charmonium production rate was larger than predicted for $J/\Psi$,
and much larger for $\Psi'$ than theoretical predictions in proton-proton (p-p)
collisions; and the disagreement between experimental measurements of 
polarization of $J/\Psi,\Psi'$ \cite{cdf00} states produced at high energy
and theoretical predictions\cite{cl96}. In addition to being an
important study of QCD, it also could provide the basis for testing the
production of Quark Gluon Plasma (QGP) in relativistic heavy ion collisions
(RHIC). We use the notation E=$\sqrt{s}$.

   At the proton-proton (p-p) energies of the Fermilab, BNL-RHIC, or the 
Large Hadron Collider the color octet
model\cite{cl96,bc96,fl96} dominates the color singlet model, as was shown in
studies of $J/\Psi$ production at E=200 GeV at BNL\cite{nlc03,cln04}. In this
color octet model heavy quark state production cross sectons require partonic 
distributions and nonperturbative matrix elements. In our present work we 
use this framework, but some of our nonperturbative matrix elements are taken 
from a recent publication\cite{lsk09} in which it was shown that some of the 
charmonium and upsilon states are mixed meson and hybrid meson states, which
followed the work in which it was shown that these states are not pure
hybrids\cite{kpr09}.  This 
is particularly important for the application of the octet model, since
the heavy quark $Q\bar{Q}$ in a hybrid are in a color octet representation.
The mixed hybrid model of Ref\cite {lsk09} can also explain the $\Psi'(2S)$ 
production anomaly, with a much larger production of $\Psi'(2S)$ in high 
energy collisions than standard model predictions\cite{cdf}, the famous 
$\rho-\pi$ puzzle, and the anomolous production of sigmas in the decay of 
$\Upsilon(3S)$ to  $\Upsilon(1S)$\cite{vogel}. In section 2 we discuss mixed 
heavy quark hybrids and the solution to these puzzles.
\clearpage

  In the present work on heavy quark state production in p-p collisions
we estimate the relative production of $\Psi'(2S)$ to $J/\Psi$ at BNL-RHIC
(200 GeV)\cite{nlc03,cln04}, and the relative production of 
$(\Upsilon(2S)+\Upsilon(3S)),\Upsilon(1S)$
states at the Large Hadron Collider (2.76 TeV)\cite{cms11}, and $(\Upsilon(nS)$
states at Fermilab (38.8 Gev)\cite{fermi91,fermi94}.

  We closely follow the formulation of Nayak and Smith\cite{ns06} in
which they calculated $J/\Psi$ and $\Psi'(2S)$ production, produced by 
unpolarized or polarized p-p collisions at $\sqrt{s}$ = 200 GeV for helicity 
$\lambda=0$ and $\lambda=1$. We generalize this by considering the
production of a heavy quark state $\Phi$, where $\Phi$ is Charmonium
$J/\Psi,\Psi'$ or Bottomonium $\Upsilon(nS)$. It is important to note that in 
Ref\cite{lsk09} the $\Psi'(2S)$ and $\Upsilon(3S)$ were found to be
approximately 50\% a heavy quark meson and 50\% a hybrid. 

\section{Review of mixed hybrid heavy quark mesons}

   The Charmonium and Upsilon (nS) states which we are studying are shown
in Figure 1
\vspace{4.5cm}

\begin{figure}[ht]
\begin{center}
\epsfig{file=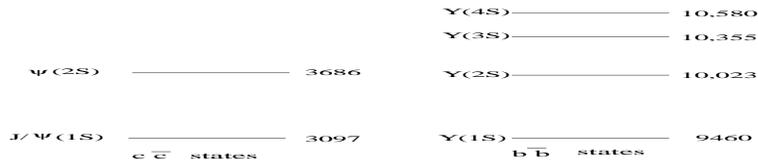,height=2cm,width=10cm}
\caption{Lowest energy Charmonium and Upsilon states}
\label{Figure 1}
\end{center}
\end{figure}

\subsection{Heavy quark meson decay puzzles}

  Note that the standard model of the $\psi'(2S)$ and $\Upsilon(3S)$ as
$c \bar{c}$ and $b \bar{b}$ mesons is not consistent with the  following
puzzles:
   
 1) The ratio of branching rarios for $c\bar{c}$ decays into hadrons (h)
given by the ratios (the wave functions at the origin canceling)
\beq
  R&=&\frac{B(\Psi'(c\bar{c})\rightarrow h)}{B(J/\Psi(c\bar{c})\rightarrow h)}
\;=\;\frac{B(\Psi'(c\bar{c})\rightarrow e^+e^-)}{B(J/\Psi(c\bar{c})\rightarrow 
e^+e^-)}\simeq 0.12 \nonumber \; ,
\eeq
the famous 12\% RULE. 

  The $\rho-\pi$ puzzle: The $\Psi'(2S)$ to $J/\Psi$ ratios for $\rho-\pi$ 
and other h decays are more than an order of magnitude too small. 
Many theorists have tried and failed to explain this puzzle.
\clearpage

2) The Sigma Decays of Upsilon States puzzle: The $\sigma$ is a broad 600 MeV 
$\pi-\pi$ resonance.
\vspace{5mm}

$\Upsilon(2s) \rightarrow \Upsilon(1S) + 2\pi$ large branching ratio. 
No $\sigma$ 
\vspace{5mm}

$\Upsilon(3s) \rightarrow \Upsilon(1S) + 2\pi$ large branching ratio to 
$\sigma$ 
\vspace{5mm}

We call this the Vogel $\Delta n=2$ Rule\cite{vogel}.  
Neither of these puzzles can be solved using standard QCD models. We
have solved them using the mixed heavy hybrid 
  
\subsection{Hybrid, mixed heavy quark hybrid mesons, and the puzzles}

  First, we used the method of QCD Sum Rules, starting with the  vector hybrid 
current ($J^{PC}=1^{--}$) for heavy quark hybrid mesons
\beq
\label{hybrid}
     J_{HH \mu} &=& \bar{q}^a_A \gamma^\nu \gamma_5 \frac{\lambda^{n}_{ab}}{2}
\tilde{G}^n_{\mu\nu}q^b_B \nonumber \; ,
\eeq
where A,B are flavor indices, a,b are color indices, and $\tilde{G}^n_{\mu\nu}$
is the gluon field operator. The QCD Sum Rule method starts with the 
two-point correlator for a heavy hybrid
\beq
\label{correlator} 
   \Pi_{HH}^{\mu\nu}(q^2) &=& i\int d^4x e^{iq \cdot x}<0|T[J_{HH}^\mu(x)
J_{HH}^\nu(0)]|0> \; .
\eeq
Using the $J_{HH \mu}$ hybrid current and the QCD Sum Rule method it was 
shown\cite{kpr09} that no low-lying charmonium state is a hybrid. 

  Then the mixed vector ($J^{PC}=1^{--}$) charmonium, hybrid charmonium current
\beq
        J^\mu &=& b J_H^\mu + \sqrt{1-b^2} J_{HH}^\mu {\rm with} \nonumber \\
          J_H^\mu &=& \bar{q}_c^a \gamma^\nu \gamma_5 q_c^a \nonumber \; ,
\eeq
{\bf where  $J_H^\mu$ is the current for a $1^{--}$ charmonium state, and 
$J_{HH}^/mu$ is the heavy charmonium hybrid current given above, was
tried\cite{lsk09}. A solution satisfying all QCD Sum Rule conditions was
found for the $\Psi'(2S)$ to be a mixed heavy hybrid, with $b\simeq -.7$
\beq
        |\Psi'(2s)>&=& -0.7 |c\bar{c}(2S)>+\sqrt{1-0.5}|c\bar{c}g(2S)> \; .
\eeq
   
  Using this method for bottom quark mesons, it was found that the 
$\Upsilon(3S)$ state was a mixed hybrid:
\beq
        |\Upsilon(3S)>&=& -0.7 |b\bar{b}(3S)>+\sqrt{1-0.5}|b\bar{b}g(3S)> \; .
\eeq  

   Therefore for both the $\Psi'(2S)$ and $\Upsilon(3S)$ b=-.7, and both
have a 50\% probability of being a hybrid and 50\% probability of being a
standard $q \bar{q}$ meson. In Ref\cite{lsk09} it was shown that this
solves the $\Psi'(2S)$ to $J/\Psi$ the $\rho-\pi$ puzzle, and the 
Vogel $\Delta n=2$ Rule.

   We now investigate how this mixed heavy hybrid meson theory affects
the production of charmonium and upsilon states via p-p collisions.

\section{Unpolarized p-p collisions} 
   The cross sections for $J/\Psi$ and $\Psi'(2S)$ production in the
color octet model are based on the cross sections obtained from the
matrix elements for quark-antiquark and gluon-gluon octet fusion to a 
hadron H, $\sigma_{q\bar{q} \rightarrow H(\lambda)}$  
and $\sigma_{gg \rightarrow H(\lambda)}$, with $\lambda$ the helicity, as 
illustrated in Figure 2.
\vspace{3 cm}

\begin{figure}[ht]
\begin{center}
\epsfig{file=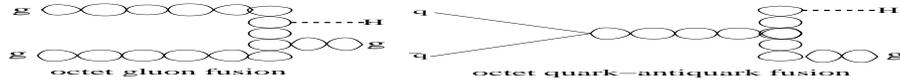,height=1.cm,width=12cm}
\caption{ Gluon and quark-antiquark color octet fusion producing hadron H}
\label{Figure.2}
\end{center}
\end{figure} 
In Ref\cite{ns06} the matrix elements used were derived by Braaten and 
Chen\cite{bc96}. The three octet matrix elements 
needed are $<O_8^\Phi(^1S_0)>$, $<O_8^\Phi(^3S_1)>$, and $<O_8^\Phi(^3P_0)>$,
with $\Phi$ either $J/\Psi$, $\Psi'(2S)$, or $\Upsilon(nS)$. Since these matrix elements
are not well known, Nyyak and Smith\cite{ns06} use three scenerios: 
1)$<O_8^\Phi(^1S_0)>$=$<O_8^\Phi(^3P_0)>/m^2$=.0087, 
2)$<O_8^\Phi(^1S_0)>$=.039 and $<O_8^\Phi(^3P_0)>$=0, 3)$<O_8^\Phi(^1S_0)>$=0
and  $<O_8^\Phi(^3P_0)>/m^2$=.01125, with $<O_8^\Phi(^3S_1)>$=.0112 in all
scenerios and all having units GeV$^3$. Note that these matrix elements 
are not used to obtain the wave functions of the heavy quark meson states.

To obtain the production cross sections one needs to multiply by the 
quark-antiquark or gluon parton distribution functions, giving
\beq
\label{ns1}
 \sigma_{pp\rightarrow \Phi(\lambda)} &=& \int_a^1  \frac{d x}{x}
f_q(x,2m)f_{\bar{q}}(a/x,2m) \sigma_{q\bar{q} \rightarrow H(\lambda)} 
\nonumber \\
&& +f_g(x,2m)f_g(a/x,2m) \sigma_{g g \rightarrow H(\lambda)} \; ,
\eeq
where $a= 4m^2/s$, with $m=1.5$  GeV for charmonium, and 5 GeV for bottomonium.
$f_g(x,2m)$, $f_q(x,2m)$ are the gluonic and quark distribution functions 
evaluated at $Q=2m$.
    First we consider unpolarized p-p collisions, using\cite{ns06} scenerio 2.
With scenerio 2 matrix elements the production cross sections\cite{bc96,ns06} 
for $\Phi$ for helicity $\lambda$ = 0 and 1 are

\beq
\label{1}
  \sigma_{pp\rightarrow \Phi(\lambda=0)} &=& A_\Phi \int_a^1 \frac{d x}{x} 
f_g(x,2m)f_g(a/x,2m) \nonumber \\
\sigma_{pp\rightarrow \Phi(\lambda=1)} &=& A_\Phi \int_a^1 \frac{d x}{x}
[f_g(x,2m)f_g(a/x,2m)+0.613((f_d(x,2m)f_{\bar{d}}(a/x,2m) \nonumber \\
    &&+f_u(x,2m)f_{\bar{u}}(a/x,2m))]
\; ,
\eeq
with $A_\Phi=\frac{5 \pi^3 \alpha_s^2}{288 m^3 s}<O_8^\Phi(^1S_0)>$,
$a= 4m^2/s$; where $m=1.5$  GeV for charmonium, and 5 GeV for bottomonium.
$f_g(x,2m)$, $f_q(x,2m)$ are the gluonic and quark distribution functions 
evaluated at $Q=2m$.

The main purpose of the present work is to explore the effects of matrix
elements for $\Psi'(2S)$ and $\Upsilon(nS)$ with n=2,3. We compare our results 
with the hybrid model to the standard model. In the standard model the states 
are (nS) quark-antiquark states, and the ratios of the matrix elements for
n greater than 1 is given by the squares of the wave functions. Note that
the basis for the octet model being used is the nonrelatavistic QCD 
model\cite{cl96,bc96,fl96}, with a model potential for the quark anti-quark
interaction giving bound states. A harmonic oscillator potential can
be used to approximately give the energies of the first few states, which is 
what is needed in the present work. For the octet matrix elements illustrated
in Figure 1, however, one must use QCD directly, and we use the results
of Refs.\cite{cl96,bc96,fl96,ns06} for these matrix elements. 

  To approximate the ratios of matrix elements in a 
nonrelativistic quark model for these heavy quark meson states we use harmonic 
oscillator wave functions\cite{merqm}, with $\Phi(1S)=2 Exp[-r/a_o]/a_o^{3/2}$,
$\Phi(2S)=\Phi(1S)(1-r/a_0)/2^{3/2}$, and $\Phi(3S)=\Phi(1S)(1-2r/3a_o
+2r^2/27a_o^2)/3^{3/2}$. Defining N1= $\int |\Phi(2S)|^2$ divided by 
$\int |\Phi(1S)|^2$ for the 2S to 1S probability, and simillarly N2 for the 
3S to 1S probability, we find N1=0.039, N2=0.0064, N3=N2/N1=.16. This is a 
very rough estimate. The cross sections for the mixed hybrid states are 
enhanced, as explained below.

   Therefore, we use $A_{\Psi'(2S)}=0.039A_{J/\Psi (1S)}$,  $A_{\Upsilon(2S)}=0.039
A_{\Upsilon (1S)}$, and $A_{\Upsilon(3S)}=0.0064 A_{\Upsilon (1S)}$ in the
standard model. On the other hand in the mixed hybrid study both $\Psi'(2S)$
and $\Upsilon(3S)$ were found to be approximately 50\% hybrids. In 
Ref\cite{lsk09} it was shown, using the external field method, that the octet 
to singlet matrix element was enhanced by a factor of $\pi^2$ compared 
to the standard model, as illustrated in Figure 3. For mixed hybrids we use
an enhancement factor of 3.0
\vspace{5mm}
\begin{figure}[ht]
\begin{center}
\epsfig{file=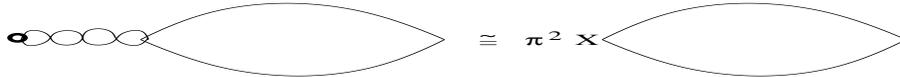,height=1.cm,width=12cm}
\caption{External field method for $\Psi'(2S)$ and $\Upsilon(3S)$ states}
\label{Figure 3}
\end{center}
\end{figure}

\newpage

  For differential cross sections we use the rapidity variable, $y$,
\beq
\label{2}
      y(x) &=& \frac{1}{2} ln (\frac{E + p_z}{E-p_z}) {\rm ;\;with\;} 
E= \sqrt{M^2 + p_z^2}  \nonumber \\
             p_z &=& \frac{\sqrt{s}}{2} (x-\frac{a}{x})  \; ,\; {\rm \;\; or}
\eeq
\beq
\label{x(y)}
      x(y) &=& 0.5 \left[\frac{m}{\sqrt{s}}(\exp{y}-\exp{(-y)})+
\sqrt{(\frac{m}{\sqrt{s}}(\exp{y}-\exp{(-y)}))^2 +4a}\right]
\eeq

   For the unpolarized proton collisions we use a polynomial fit to the parton 
distributions of Ref.\cite{CTEQ6}. Because of the wide range of vaues, in 
order to obtain a good polynomial fit to the parton distributions we limit 
the range of rapidity to $-1. < y <1.$

   For Q=3 GeV, with m=Charmonium mass = 1.5 GeV, from Eq(\ref{x(y)}), x has a
range about 0.028 to 0.032, and a/x 0.008 to 0.015.  We have derived the 
following expressions for the gluon (g), u and d quark, and antiquark 
distribution functions using QTEQ6 for Q=3 GeV, fitting the range x=0.008 to 
.004, which is needed for $\sqrt{s}$=200 GeV
\beq
\label{3}
      f_g(x) & \simeq & 1334.21 - 67056.5 x + 887962.0 x^2  \nonumber \\
      f_d(x) & \simeq &72.956 - 3281.1 x + 42247.6 x^2 \nonumber \\
      f_u(x) & \simeq & 82.33 - 3582.36 x + 45867.3 x^2 \nonumber \\
  f_{\bar{u}}(x) & \simeq &55.98 - 2722.04 x + 35641.2 x^2 \\  
  f_{\bar{d}}(x) & \simeq &57.44 - 2757.05 x + 36030.5 x^2 \nonumber \; .
\eeq

  For Q=10 GeV, m=Bottomonium mass=5 GeV, from Eq(\ref{x(y)}), x has a
range about 0.05 to 0.08, and a/x 0.03 to 0.05. We have derived the 
following expressions for the gluon (g), u and d quark, and antiquark 
distribution functions using QTEQ6 for Q=10 GeV, fitting the range x=0.03 to 
.08, which is needed for $\sqrt{s}$=38.8 GeV and 2.76 TeV.
\beq
\label{4}
      f_g(x) & \simeq & 275.14 - 6167.6 x + 36871.3 x^2 \nonumber \\
      f_d(x) & \simeq & 26.96 - 527.14 x + 3119.13 x^2 \nonumber \\
      f_u(x) & \simeq & 32.92 - 604.38 x + 3530.1 x^2 \nonumber \\
  f_{\bar{u}}(x) & \simeq & 16.64 - 377.53 x + 2336.86 x^2 \\  
  f_{\bar{d}}(x) & \simeq & 17.81 - 390.64 x + 2392.46 x^2 \nonumber \; .
\eeq

The differential rapidity distribution for $\lambda=0$ is given by
\beq
\label{5}
      \frac{d \sigma_{pp\rightarrow \Phi(\lambda=0)}}{dy} &=& 
     A_\Phi \frac{1}{x(y)} f_g(x(y),2m)f_g(a/x(y),2m) \frac{dx}{dy} \; ,
\eeq

while for $\lambda$=1
\beq
\label{6}
\frac{d \sigma_{pp\rightarrow \Phi(\lambda=1)}}{dy} &=& A_\Phi \frac{1}{x(y)}
[f_g(x(y),2m)f_g(a/x(y),2m)+0.613(f_d(x(y),2m)f_{\bar{d}}(a/x(y),2m) 
\nonumber \\
    &&+f_u(x(y),2m)f_{\bar{u}}(a/x(y),2m)]\frac{dx}{dy} \; .
\eeq

\clearpage

 \subsection{Charmonium  Production Via Unpolarized p-p
Collisions at E=$\sqrt{s}$= 200 Gev at BNL-RHIC}

    First we consider unpolarized p-p collisions for $\sqrt{s}=200 GeV$  
corresponding to BNL energy. We use scenerio 2\cite{ns06}, with the 
nonperturbative matrix elements given above. Therefore, $A_\Phi=
\frac{5 \pi^3 \alpha_s^2}{288 m^3 s} <O_8^\Phi(^1S_0)>$ =$7.9 \times 10^{-4}$nb 
for $\Phi$=$J/\Psi$ and $2.13 \times  10^{-5}$nb for $\Upsilon(1S)$ heavy 
quark states; $a= 4m^2/s = 2.25 \times 10^{-4}$ for Charmonium and 
$2.5 \times 10^{-3}$ for Bottomium.

For  $\sqrt{s}=200 GeV$ 
\beq
\label{7}
   x(y) &=& 0.5 \left[\frac{m}{200}(\exp{y}-\exp{(-y)})+\sqrt{(\frac{m}{200}
(\exp{y}-\exp{(-y)}))^2 +4a}\right] \nonumber \\
  \frac{d x(y)}{d y} &=&\frac{M}{400}(\exp{y}+\exp{(-y)})\left[1. + 
\frac{\frac{M}{200}(\exp{y}+\exp{(-y)})}{\sqrt{(\frac{M}{200} 
(\exp{y}-\exp{(-y)}))^2 +4a}}\right] \; .
\eeq

   Using Eqs(\ref{5},\ref{6},\ref{7}), with the parton distribution functions 
given in Eq(\ref{3}), we find d$\sigma$/dy for Q=3 GeV, $\lambda=0$ and 
$\lambda=1$ the results for $J/\Psi$ shown in Figure 4.

\begin{figure}[ht]
\begin{center}
\epsfig{file=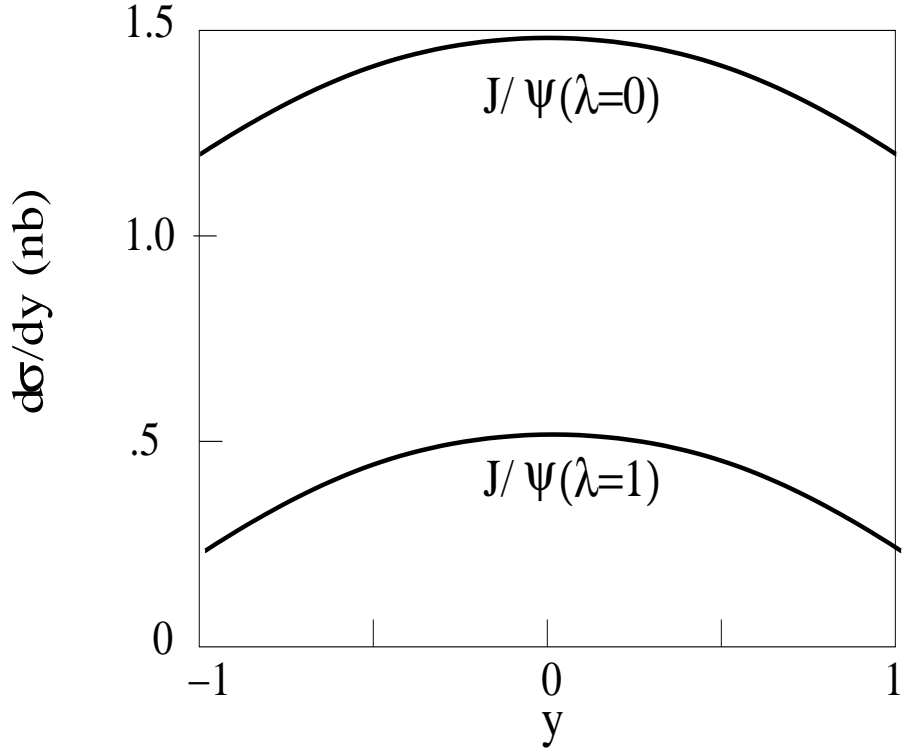,height=10 cm,width=12cm}
\caption{d$\sigma$/dy for Q=3 GeV, E=200 GeV unpolarized p-p collisions 
producing $J/\Psi$
with $\lambda=0$, $\lambda=1$}
\label{Figure 4}
\end{center}
\end{figure}

Note that the shape of d$\sigma$/dy is consistent with the BNL-RHIC-PHENIX
detector rapidity distribution\cite{cln04}.
\clearpage

For $\Psi'(2S)$ the results are shown in Figure 5.
\vspace{6cm}

\begin{figure}[ht]
\begin{center}
\epsfig{file=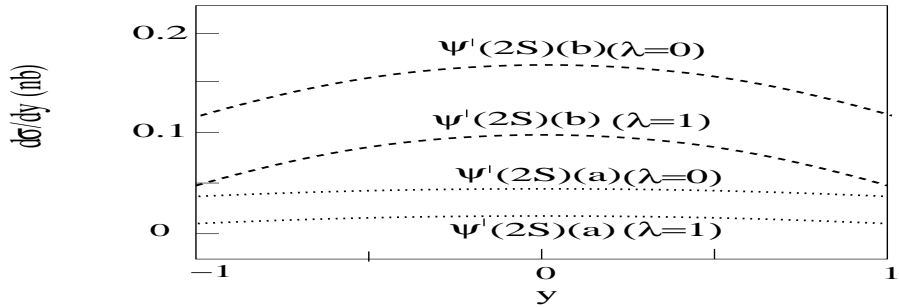,height=4cm,width=12cm}
\caption{d$\sigma$/dy for Q= 3 GeV, E=200 GeV unpolarized p-p  collisions 
producing$\Psi'(2S)$
 with $\lambda=1$,$\lambda=0$}
\label{Figure 5}
\end{center}
\end{figure}

The results for d$\sigma$/dy shown in Figure 4.  labeled $\Psi'(2S)(a)$ are
 obtained by using for the standard nonperturbative matrix element=0.039 
times the matrix elements for $J/\Psi$ production; while the results labeled 
$\Psi'(2S)(b)$ is obtained by using
the matrix element derived using the result that the $\Psi'(2S)$ is 
approximately 50\% a hybrid with the enhancement is at least a factor of $\pi$,
as discussed above.

\subsection{Upsilon Production Via Unpolarized p-p
Collisions at E=$\sqrt{s}$= 38.8 Gev at Fermilab}

Since the $\Upsilon(nS)$ states have not been resolved at BNL-RHIC at the 
present time, we consider $\Upsilon(nS)$ state production at 38.8, which has
been measured at Fermilab\cite{fermi91,fermi94}

 For Q=10 GeV, using the parton distributions given in Eq(\ref{4}) and 
Eqs(\ref{5},\ref{6}) for helicity $\lambda=0$, $\lambda=1$, with
$A_{\Upsilon}$ =$5.66 \times 10^{-4}$nb and $a=6.64 \times 10^{-2}$, we obtain 
$d\sigma/dy$ for $\Upsilon(nS)$ production.

\clearpage

 The results for $\Upsilon(1S)$, $\Upsilon(2S)$ are shown in Figure 6, 
and for $\Upsilon(3S)$ in Figure 7.
\vspace{7.0cm}

\begin{figure}[ht]
\begin{center}
\epsfig{file=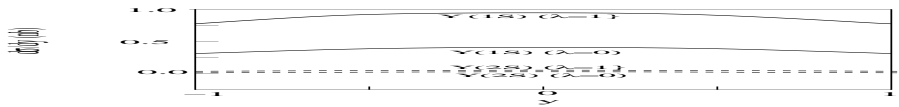,height=1.3cm,width=12cm}
\caption{d$\sigma$/dy for Q= 10 GeV, E=38.8 GeV unpolarized p-p collisions 
producing $\Upsilon(1S)$, $\Upsilon(2S)$ with $\lambda=0$, $\lambda=1$}
\label{Figure 6}
\end{center}
\end{figure}

\vspace{6cm}

\begin{figure}[ht]
\begin{center}
\epsfig{file=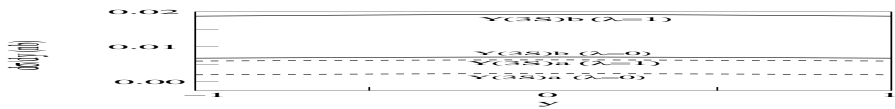,height=1.3cm,width=12cm}
\caption{d$\sigma$/dy for Q= 10 GeV, E=38.8 GeV unpolarized p-p collisions 
producing $\Upsilon(3S)$ with $\lambda=0$, $\lambda=1$ }
\label{Figure 7}
\end{center}
\end{figure}

\clearpage

   In Figure 6 $d\sigma/dy$ for $\Upsilon(1S)$ and $\Upsilon(2S)$ are
obtained using the standard model for the matrix elements. In figure 7
the results for $d\sigma/dy$ for the standard model are labelled
$\Upsilon(3S)$a, while the results for the hybrid model are labelled
$\Upsilon(3S)$b 

 It should be noted that the ratios of $d\sigma/dy$ for $\Psi'(2S)$, shown
in Figure 5 and $\Upsilon(3S)$, shown in Figure 7 for the hybrid theory vs. 
the standard are our most significant results, as there are uncertainties in 
the absolute magnitudes and shapes of $d\sigma/dy$ on the scenerios, as
well as the magnitudes of the matrix elements. This is discussed in the
following subsection.

\subsection{Dependence of $d \sigma /dy$ on scenerios}

As stated above, our calculations make use of scenerio 2 of Ref\cite{ns06}.
It is important for us to point out that our main objective in the present
work is to derive the relative magnitudes of the $\Psi'(2S)$ and $\Upsilon(3S)$
cross sections in our mixed heavy hybrid theory, vs standard quark models
for these states. To ilustrate this, we show Fig. 4 from Nayak and Smith'
2006 publication\cite{ns06}. Scenerios 1, 2, 3 are
defined by  1 $<O_8^\Phi(^3P_0)>$=  $<O_8^\Phi(^1S_0)>$= $.0087 GeV^3$, 
2 $<O_8^\Phi(^3P_0)>$=0 and $<O_8^\Phi(^1S_0)>$=$.039 GeV^3$, and 
3 $<O_8^\Phi(^3P_0)>$= $.01123 GeV^3$ and $<O_8^\Phi(^1S_0)>$= 0. See
Ref\cite{ns06} for the form of Eq(1) with the three scenerios.
\vspace{4cm}

\begin{figure}[ht]
\begin{center}
\epsfig{file=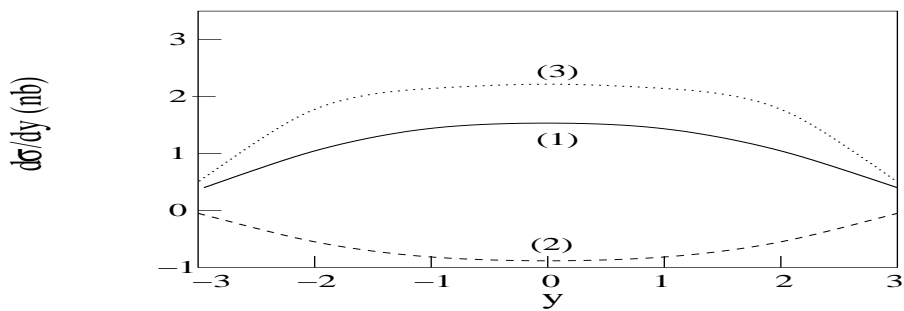,height=4.0cm,width=12cm}
\caption{d$\sigma$/dy for Q= 3 GeV, E=200 GeV polarized p-p collisions 
with $\lambda=1$ producing $J/\Psi$. Scenerios 1, 2, 3 are shown with
solid, dashed, and dotted curves, from Ref\cite{ns06}} 
\label{Figure 8}
\end{center}
\end{figure}

  As one can see from Figure 8, the shapes and magnitudes of 
$d \sigma /dy$ depends on the scenerios, but this is not a problem in the 
present work, since we are mainly interested in ratios of cross sections; 
and our main problem with the shape (see Figs. 9 and 10) are for polarized 
$\Psi$ production as in Ref\cite{ns06}.

\clearpage

\section{Polarized p-p collisions at 200 Gev at BNL-RHIC} 

For polarized p-p collisions the equations for  
$\frac{d \sigma_{pp\rightarrow \Phi(\lambda=0)}}{dy}$ and
$ \frac{d \sigma_{pp\rightarrow \Phi(\lambda=1)}}{dy}$ are the same as 
Eqs(\ref{5},\ref{6}) with the parton distribution functions $fg$ and $fq$
given in Eqs(\ref{3},\ref{4}) replaced by $\Delta fg$ and $\Delta fq$,
the parton distribution functions for longitudinally polarized p-p collisions.

A fit to the parton distribution functions for polarized p-p collisions
for Q=3 GeV obtained from CTEQ6\cite{CTEQ6} in the x range needed for
$\sqrt{s}$=200 GeV is
\beq
\label{8}
\Delta f_g(x) &\simeq & 15.99-700.34 x+13885.4 x^2-97888. x^3 \nonumber \\
\Delta f_d(x) & \simeq & -5.378.+205.60 x-4032.77 x^2+28371. x^3 \nonumber \\
\Delta f_u(x) & \simeq & 8.44-292.19 x+5675.16 x^2-39722. x^3 \nonumber \\
\Delta f_{\bar{u}}(x) & \simeq &-1.447 +64.67 x-1268.24 x^2+8878.32 x^3  \\  
  \Delta f_{\bar{d}}(x) &=&\Delta f_{\bar{u}}(x) \nonumber \; ,
\eeq
and for Q=10 GeV, which we do not use in the present work, as the $\Upsilon(nS)$
are not resolved at BNL-RHIC,

\beq
\label{9}
\Delta f_g10(x) & \simeq & 28.98-1435.47 x+29533.5 x^2-211440. x^3 \nonumber \\
\Delta f_d10(x) & \simeq &-6.074+241.57 x-4762.04 x^2+33604.4 x^3 \nonumber \\
\Delta f_u10(x) & \simeq & 9.88-348.632 x+6729.49 x^2-47058. x^3  \nonumber \\
\Delta f_{\bar{u}}10(x) & \simeq & -1.552+75.731 x-1531.97 x^2+10896.6 x^3  \\  
\Delta f_{\bar{d}}10(x) &=& \Delta f_{\bar{u}}10(x) \nonumber \; .
\eeq

The differential rapidity distribution for polarized p-p collisions are
\beq
\label{Deltasig0}
      \frac{d \Delta \sigma_{pp\rightarrow \Phi(\lambda=0)}}{dy} &=& 
  - A_\Phi \frac{1}{x(y)}\Delta f_g(x(y),2m)\Delta f_g(a/x(y),2m) \frac{dx}{dy} 
\; ,
\eeq

\beq
\label{Deltasig1}
\frac{d \Delta \sigma_{pp\rightarrow \Phi(\lambda=1)}}{dy} &=& -A_\Phi 
\frac{1}{x}[\Delta f_g(x(y),2m) \Delta f_g(a/x(y),2m)-0.613(\Delta f_d(x(y),2m) 
\nonumber \\
 &&\Delta f_{\bar{d}}(a/x(y),2m) +\Delta f_u(x(y),2m)
\Delta f_{\bar{u}}(a/x(y),2m)]\frac{dx}{dy} \; .
\eeq

For polarized p-p collisions, Q=3 GeV,  the results for d$ \Delta \sigma$/dy for
$J/\Psi$ production using the standard model are shown in Figure 9; 
while for $\Psi'(2S)$ the results are shown in Figure 10.  As above, the 
curves labelled $\Psi'(2S)a$ and are the standard model results, while that 
labelled $\Psi'(2S)b$ are the results for a mixed hybrid.
The enhancement from active glue is once more quite evident.
Since $\Upsilon(nS)$ states have not been resolved at BNL-RHIC, where polarized
p-p collisions were measured, we do not calculate d$ \Delta \sigma$/dy for 
$\Upsilon(nS)$ states.

\clearpage

\begin{figure}[ht]
\begin{center}
\epsfig{file=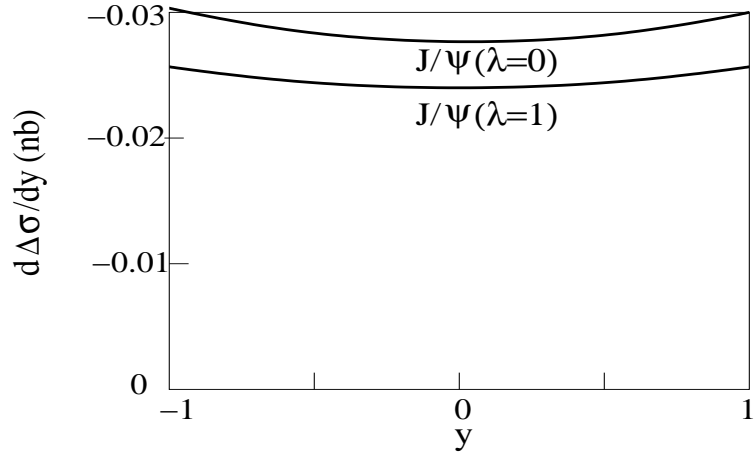,height=6cm,width=10cm}
\caption{d$ \Delta \sigma$/dy for Q=3 GeV, E=200 GeV polarized p-p collisions 
producing $J/\Psi$, with $\lambda = 0$, $\lambda = 1$}
\label{Figure 9}
\end{center}
\end{figure}

\begin{figure}[ht]
\begin{center}
\epsfig{file=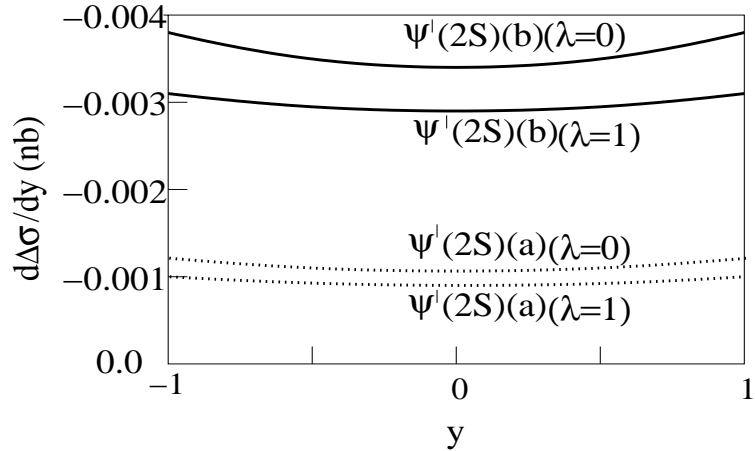,height=6cm,width=10cm}
\caption{d$ \Delta \sigma$/dy for Q= 3 GeV, E=200 GeV polarized p-p  collisions 
producing $\Psi'(2S)$
with $\lambda=0$, $\lambda=1$}
\label{Figure 10}
\end{center}
\end{figure}

Once again, we stress that it is the ratios of $d \Delta \sigma/dy$ that is 
most significant, as there is uncertainty both the absolute magnitudes and
shapes. Note that d$ \Delta \sigma$/dy is expected to have a maximum at y=0. 
As in N-S for polarized $J/\Psi$ production, $d \Delta \sigma/dy <0$ but there 
is a minimum at y=0 in absolute value for our results with
scenerio 2. (See preceeding subsection on dependence on scenerios.)

\clearpage

 \section{Upsilon Production Via Unpolarized p-p
Collisions at 2.76 Tev at LHC-CMS and 38.8 GeV at Fermilab}

  The cross sections for $\Upsilon(nS)$ state production in p-p collisions
have been measured at 2.76 TeV at the LHC-CMS\cite{cms11} and at 38.8 GeV
at Fermilab\cite{fermi91}. In this subsection we calculate the cross sections
for $\Upsilon(nS)$ production, with n= 1, 2, 3. Then we use our theory that
$\Upsilon(3S)$ is a hybrid to estimate the ratios of cross section. Since
we are using scenerio 2 with $<O_8^\Phi(^3P_0)>$=0, the $\lambda=0$ helicity
dominates the cross section\cite{ns06}, and we drop the $\lambda=1$ terms.  
From Eq(\ref{1}), for $\lambda=0$, the cross section is determined from
\beq
  \sigma_{pp\rightarrow \Phi(\lambda=0)} &=& A_\Phi \int_a^1 \frac{d x}{x} 
fg(x,2m)fg(a/x,2m) \nonumber \; .
\eeq

We use this to estimate the ratios of the $\Upsilon(2S)$ and $\Upsilon(3S)$
production cross sections to the $\Upsilon(1S)$ production cross section for
2.76 TeV and 38.8 GeV.

\subsection{Upsilon Production Via Unpolarized p-p Collisions at 2.76 Tev}
 
 For $\sqrt{s}$=2.76 TeV, for $\Upsilon(nS)$ production,
\beq
\label{A2.76}
      a&=&1.31 \times 10^{-5} \nonumber \\
      A_\Upsilon &=& 1.12 \times 10^{-7} \; .
\eeq      

  The cross sections for $\Upsilon(nS)$ state production in p-p collisions
have been measured at 2.76 TeV at the LHC-CMS\cite{cms11} and at 38.8 GeV
at Fermilab\cite{fermi91}. In this subsection we calculate the cross sections
for $\Upsilon(nS)$ production, with n= 1, 2, 3. Then we use our theory that
$\Upsilon(3S)$ is a hybrid to estimate the ratios of cross section. Since
we are using scenerio 2 with $<O_8^\Phi(^3P_0)>$=0, the $\lambda=0$ helicity
dominates the cross section\cite{ns06}, and we drop the $\lambda=1$ terms.  
From Eq(\ref{1}), for $\lambda=0$, the cross section is determined from
\beq
  \sigma_{pp\rightarrow \Phi(\lambda=0)} &=& A_\Phi \int_a^1 \frac{d x}{x} 
fg(x,2m)fg(a/x,2m) \nonumber \; .
\eeq

From this we find 
\beq
\label{sigupsilon}
      \sigma_{pp\rightarrow \Upsilon(1S)}&\simeq & 0.85 {\rm \;\; nb}
\eeq

What is significant for the present work are the ratios of the (1S), (2S), (3S)
state production. In the preceeding sections we used estimates from the wave
functions to find these ratios for the differential cross section. In the
present section we make use of experimental results from the Fermilab
experiment\cite{fermi91} for the ratio of $\sigma(\Upsilon(2S))/
\sigma(\Upsilon(1S))$, since the 2S and 1S states are given by the standard
model, and $\sigma(\Upsilon(3S))/\sigma(\Upsilon(1S)) =N3 \times 
\sigma(\Upsilon(2S))/\sigma(\Upsilon(1S))$, as discussed 
in Section 2. Therefore, our estimate of the standard model is

\beq
\label{2S3S/1S}
      \sigma(\Upsilon(2S))/\sigma(\Upsilon(1S))&\simeq& 0.27 {\rm \;\;standard}
\nonumber \\
      \sigma(\Upsilon(3S))/\sigma(\Upsilon(1S))&\simeq& 0.04 
{\rm \;\;standard,\;\; giving}
\nonumber \\
 \frac{\sigma(\Upsilon(2S))+\sigma(\Upsilon(3S))}
{\sigma(\Upsilon(1S))} &\simeq& 0.31 {\rm \;\;standard}
\eeq

On the other hand, in our mixed hybrid theory with the $\Upsilon(3S)$ about 50\%
hybrid\cite{lsk09}, we would expect a factor of $\pi^2/4$ in the matrix element,
and therefore a factor of about 2.45 for the $\Upsilon(3S)$ cross section
compared to the standard model. 

 This results in our estimate
\beq
        \frac{\sigma(\Upsilon(2S))+\sigma(\Upsilon(3S))}{\sigma(\Upsilon(1S))}
&=& 0.52 \; ,
\eeq
Compared to the LHC-CMS result\cite{cms11} that this ratio is 
$0.78^{.16}_{-.14} \pm .02$. while in the standard model it would be about 0.31.

\subsection{Upsilon Production Via Unpolarized p-p Collisions at 38.8
GeV}

Our study of the 38.8 Upsilon production is similar to the preseeding one
for the LHC-CMS 2.76 TeV experiments. Our result for the 
$\sigma(\Upsilon(3S))/\sigma(\Upsilon(1S))$ expected at 38.8 GeV in the
standard model, see Eq(\ref{2S3S/1S}, compared to our mixed hybrid theory:
\beq
   \frac{\sigma(\Upsilon(3S))}{\sigma(\Upsilon(1S))}&\simeq& 0.04 
{\rm \;\;standard} \nonumber \\
   \frac{\sigma(\Upsilon(3S))}{\sigma(\Upsilon(1S))}&\simeq& 0.147-0.22
 {\rm \;\;hybrid} 
\eeq
compared to the experimental result\cite{fermi91} of about 0.12 to 0.16 

\section{Conclusions}

   We have applied the mixed hybrid theory for heavy quark states and 
predict that the cross sections for production of the charmonium $\Psi'(2S)$
state in 200 GeV p-p collisions and bottomonium $\Upsilon(3S)$ states in 38.8
 GeV p-p collisions are much larger than the standard model. We have also 
estimated ratio of cross sections for 2.76 TeV and 38.8 GeV experiments, and 
our prediction for the $\Upsilon(3S)$ production cross section is larger 
than the standard model, and closer to the experimental values. 

  Because of the importance of gluonic production in processes in 
a Quark Gluon Plasma, this could lead to a test of the creation of 
QGP in RHIC. In order to treat the production of heavy mixed hybrid 
states via the QGP, however, we must redo the QCD calculation using 
finite temperature field theory, as the properties of both the standard 
and hybrid components of the $\Psi'(2S)$ and $\Upsilon(3S)$, as well
as the properties of the $J/\Psi$ and the other $\Upsilon(nS)$ states,
are modified by the temperature during the QCD phase transition. This 
is a project for future research.
\vspace{3mm}

\Large{{\bf Acknowledgements}}\\
\normalsize
This work was supported in part by a grant from the Pittsburgh Foundation,
and in part by the DOE contracts W-7405-ENG-36 and DE-FG02-97ER41014.

\end{document}